\documentclass[aps,showpacs,preprint,preprintnumbers,amsmath,amssymb]{revtex4}
\usepackage{bm}% bold math

\usepackage[T1]{fontenc}
\usepackage{amssymb,amsmath,amsthm}
\usepackage{amsfonts}
\usepackage{graphicx, pstricks}

\begin{document}
\def\vec#1{\mathbf{#1}}
\def\ket#1{|#1\rangle}
\def\bra#1{\langle#1|}
\def\ketbra#1{|#1\rangle\langle#1|}
\def\braket#1{\langle#1|#1\rangle}
\def\idmat{\mathbf{1}}
\def\caln{\mathcal{N}}
\def\calc{\mathcal{C}}
\def\rhon{\rho_{\mathcal{N}}}
\def\rhoc{\rho_{\mathcal{C}}}
\def\tr{\mathrm{tr}}
\def\bfu{\mathbf{u}}
\def\bfmu{\mbox{\boldmath$\mu$}}

\newcommand{\ri}{{\rm i}}
\newcommand{\re}{{\rm e}}
\newcommand{\bb}{{\bf b}}
\newcommand{\bc}{{\bf c}}
\newcommand{\bx}{{\bf x}}
\newcommand{\bz}{{\bf z}}
\newcommand{\by}{{\bf y}}
\newcommand{\bv}{{\bf v}}
\newcommand{\bd}{{\bf d}}
\newcommand{\br}{{\bf r}}
\newcommand{\bk}{{\bf k}}
\newcommand{\bA}{{\bf A}}
\newcommand{\bE}{{\bf E}}
\newcommand{\bF}{{\bf F}}
\newcommand{\bR}{{\bf R}}
\newcommand{\bM}{{\bf M}}
\newcommand{\bn}{{\bf n}}
\newcommand{\bs}{{\bf s}}
\newcommand{\tbs}{\tilde{\bf s}}
\newcommand{\rSi}{{\rm Si}}
\newcommand{\cL}{{\cal L}_x}
\newcommand{\dB}{d_{\rm Bures}}
\newcommand{\beps}{\mbox{\boldmath{$\epsilon$}}}
\newcommand{\bthe}{\mbox{\boldmath{$\theta$}}}
\newcommand{\blam}{\mbox{\boldmath{$\lambda$}}}
\newcommand{\rg}{{\rm g}}
\newcommand{\xmax}{x_{\rm max}}
\newcommand{\ra}{{\rm a}}
\newcommand{\rx}{{\rm x}}
\newcommand{\rs}{{\rm s}}
\newcommand{\rP}{{\rm P}}
\newcommand{\up}{\uparrow}
\newcommand{\down}{\downarrow}
\newcommand{\hc}{H_{\rm cond}}
\newcommand{\kb}{k_{\rm B}}
\newcommand{\cI}{{\cal I}}
\newcommand{\tit}{\tilde{t}}
\newcommand{\cE}{{\cal E}}
\newcommand{\cC}{{\cal C}}
\newcommand{\Ubs}{U_{\rm BS}}
\newcommand{\qq}{{\bf ???}}
\newcommand*{\etal}{\textit{et al.}}

\newcommand{\be}{\begin{equation}}
\newcommand{\ee}{\end{equation}}
\newcommand{\bfg}{\begin{figure}}
\newcommand{\efg}{\end{figure}}
\newcommand{\Itwo}{\mathbb{1}_2}
\newcommand{\I}{\mathcal{I}}
\newcommand{\al}{\alpha}

\sloppy

\title{Parameter estimation with mixed quantum states} 
\author{Daniel Braun$^{1,2}$}
\affiliation{$^1$ Universit\'e de Toulouse, UPS, Laboratoire
de Physique Th\'eorique (IRSAMC), F-31062 Toulouse, France}
\affiliation{$^2$ CNRS, LPT (IRSAMC), F-31062 Toulouse, France}
%\centerline{\today}
\begin{abstract}
We consider quantum enhanced measurements with initially mixed states.  We
show very generally that for any linear propagation of
the initial state that depends smoothly on the parameter to be estimated,
the sensitivity is bound by the maximal sensitivity that can be
achieved for any of the pure states from which the initial density matrix is
mixed.  This provides a very general proof that purely classical correlations
cannot improve the sensitivity of parameter estimation schemes in quantum
enhanced measurement schemes.  
\end{abstract}
%\pacs{03.67.-a, 03.67.Lx, 03.67.Mn }
\maketitle

Quantum enhanced measurements aim at exploiting quantum mechanical effects
for improving the sensitivity of measurements of classical parameters of a
system.  The most-well known examples are the improvement of the measurement
of phase shifts in an interferometer by using squeezed light
\cite{Caves81}, or the idea of using NOON states for super-resolution
\cite{Boto00}.  In 1994 Braunstein and Caves developed a general theory of
quantum enhanced measurements, based on a generalization to the quantum
world of the celebrated
Cram\'er-Rao bound in classical parameter estimation theory
\cite{Braunstein94,Cramer46}. Indeed, even classically the problem of the
optimal strategy of determining the value of a parameter $x$ that
parametrizes a probability distribution, and the best achievable sensitivity,
are of fundamental importance in many branches of science.  In
quantum mechanics, one would like to get the best possible estimate of a
variable 
$x$ that parametrizes a general quantum state $\rho(x)$, based on arbitrary
measurements, repeated eventually $N$ times in an always identically
prepared state.  In the above example of an interferometer,
$\rho(x)$ would be the state at the output of the interferometer, and $x$
the phase shift we want to measure.  The latter might itself depend on a
number of 
interesting parameters, such as the length of an arm of the interferometer
in the case of a Mach-Zehnder interferometer, the absorption index of a
probe brought into one of the arms, or the magnetic field acting on a spin in
the case of Ramsey interferometry. \\

Braunstein and Caves showed \cite{Braunstein94} that the
best achievable sensitivity for measuring $x$ in the state $\rho(x)$
is given by 
\begin{equation} \label{dxmin}
\langle \delta x^2\rangle_{\rho(x)}^{1/2}\ge
\frac{1}{\sqrt{N}\left(\frac{ds^2}{dx^2}\right)^{1/2}}\equiv 
\delta x_{\rm min}\left.\right|_{\rho(x)}\,.
\end{equation}
Here, $ds^2=4d_{\rm Bures}^2(\rho(x),\rho(x)+d\rho)$ is given by the so-called
Bures distance, which will be defined precisely below, and $d\rho$ denotes the
infinitesimal change of $\rho(x)$ when $x$ varies about its initial value
for which we want to know the best achievable sensitivity.  It is important
to note 
that the bound (\ref{dxmin}) can be saturated in the limit of large $N$. 

Based on (\ref{dxmin}), a number of important results were obtained.
Notably, it was shown that for an initial product state
$|\psi\rangle=\otimes_{i=1}^K\ket{\phi}_i$ of $K$ sub-systems (i.e.~$K$ ``quantum
resources''), unitary evolution with a hamiltonian $H(x)=x\sum_{i=1}^K
h_i$,  and arbitrary final projective measurements, the best achievable
sensitivity is  
\begin{equation} \label{dxminUs}
\delta x_{\rm min}\left.\right|_{|\psi(x)\rangle\langle\psi(x)|}=\frac{1}{\sqrt{NK}(\Lambda-\lambda)}\,,
\end{equation}
where $|\psi(x)\rangle=\exp(-iH(x))|\psi\rangle$, and $\Lambda$ and
$\lambda$ are the largest and smallest eigenvalue of $h_i$, respectively
(the $h_i$ are 
taken identical for all subsystems).  On the other hand, if the 
initial state is fully entangled over all $K$ subsystems, the same unitary
evolution allows to achieve a sensitivity 
\begin{equation} \label{dxminU}
\delta x_{\rm
  min}\left.\right|_{|\psi(x)\rangle\langle\psi(x)|}=
  \frac{1}{\sqrt{N}K(\Lambda-\lambda)}\,,  
\end{equation}
i.e.~an improvement by a factor $1/\sqrt{K}$ \cite{Giovannetti06}. 

The development of quantum enhanced measurements has focused so far on
pure initial states. These 
are either entangled or product states. The decisive degree of freedom
is then the number of subsystems over which entanglement extends.  In the
case of mixed states, there is more freedom, as classical 
correlations and entanglement might coexist.  The question thus arises what
can be 
achieved with general initial mixed states, and in particular whether
classical correlations may improve quantum enhanced measurements.  In this
Brief Report we
prove a no--go theorem, which essentially says that with a mixed state one
cannot achieve better sensitivity than with any of the pure states from
which it is mixed. 
In order to prove this result, we first show a little Lemma.

{\em Lemma.} The squared Bures distance is joint-convex, i.e.~for two
arbitrary states $\rho$, $\sigma$ we have 
\begin{eqnarray} \label{lem}
ds_{\rm Bures}^2(a\rho_1+(1-a)\rho_2,a\sigma_1+(1-a)\sigma_2)
&\le& a\, ds_{\rm
  Bures}^2(\rho_1,\sigma_1)+(1-a) ds_{\rm  Bures}^2(\rho_2,\sigma_2)\,.
\end{eqnarray}
\begin{proof} The squared Bures distance is defined as
\begin{equation} \label{db2}
ds_{\rm Bures}^2(\rho,\sigma)\equiv 2\left(1-\sqrt{F(\rho,\sigma)}\right)\,,
\end{equation}
where the fidelity $F(\rho,\sigma)$ is given by
$F(\rho,\sigma)=||\rho^{1/2}\sigma^{1/2}||^2_1$, and $||A||_1\equiv {\rm
  tr}\sqrt{AA^\dagger}$ denotes the trace norm \cite{Miszczak09}. It is
known that $\sqrt{F(\rho,\sigma)}$ is joint-concave \cite{Miszczak09}, i.e.~we
have 
\begin{eqnarray} \label{sF}
\sqrt{F(a\rho_1+(1-a)\rho_2,a\sigma_1+(1-a)\sigma_2)}
&\ge& a\sqrt{F(\rho_1,\sigma_1)}+(1-a)\sqrt{F(\rho_2,\sigma_2)}\,.
\end{eqnarray}
This immediately gives the joint-convexity of the squared Bures distance,
\begin{eqnarray}
ds_{\rm
  Bures}^2(a\rho_1+(1-a)\rho_2,a\sigma_1+(1-a)\sigma_2)
&=&2\left(1-\sqrt{F(a\rho_1+(1-a)\rho_2,a\sigma_1+(1-a)\sigma_2)}\right)\nonumber\\
&\le&
  2\left(1-\left(a\sqrt{F(\rho_1,\sigma_1)}+(1-a)\sqrt{F(\rho_2,\sigma_2)}\right)\right)\nonumber\\ 
&\le&a\,
  2\left(1-\sqrt{F(\rho_1,\sigma_1)}\right)+(1-a)\,2\left(1-\sqrt{F(\rho_2,\sigma_2)}\right)\nonumber\\ 
&\le&a\,d_{\rm Bures}^2(\rho_1,\sigma_1)+(1-a)\,d_{\rm
  Bures}^2(\rho_2,\sigma_2)\,,\nonumber
\end{eqnarray}
which proves the Lemma.
\end{proof}

With this Lemma, it is straightforward to show the following theorem:

{\bf Theorem.} Let $\rho_0=\sum_i p_i|\psi_i\rangle\langle\psi_i|$ be an
arbitrary mixed state ($0\le p_i\le 1$, $\sum_i p_i=1$), and $\rho(x)={\cal
  L}_x[\rho_0]$ the state generated from $\rho_0$ by an arbitrary linear map
${\cal L}_x$ which depends smoothly on the parameter $x$.  Then the
sensitivity for measuring $x$ in the state $\rho(x)$ is bounded by the 
best achievable sensitivity in the state obtained by propagating any of the
pure states $|\psi_i\rangle$ 
with the same map, i.e.
\begin{equation} \label{thm}
\delta x_{\rm min}\left.\right|_{\rho(x)}\ge \delta x_{\rm
  min}\left.\right|_{{\cal L}_x[|\psi_0\rangle\langle\psi_0|]}\,,
\end{equation}
where $|\psi_0\rangle$ is the pure state in the set of all $\{|\psi_i\rangle\}$
which leads to the best sensitivity, i.e.~${\rm
  min}_{\{|\psi_i\rangle\}}\delta x_{\rm min}\left.\right|_{{\cal
    L}_x[|\psi_i\rangle\langle\psi_i|]}=\delta x_{\rm
  min}\left.\right|_{{\cal L}_x[|\psi_0\rangle\langle\psi_0|]}$. 

\begin{proof} Due to the linearity of ${\cal L}_x[\rho]$, we have 
\begin{equation} \label{rhox}
\rho(x)={\cal L}_x[\rho_0]=\sum_i p_i {\cal L}_x[\ketbra{\psi_i}]\,,
\end{equation}
and thus, for smooth $x$-dependence of ${\cal L}_x$,
\begin{equation} \label{drho}
d\rho=\frac{d\rho}{dx}dx=\sum_i p_i dx\cL'[\ketbra{\psi_i}]\,,
\end{equation}
where $\cL'$ denotes the derivative of $\cL$ with respect to $x$.
Therefore,
\begin{eqnarray} \label{db2x}
\dB^2(\rho(x),\rho(x)+d\rho)&=&\dB^2\left(\sum_i p_i \cL[\ketbra{\psi_i}],\sum_i
p_i \left(\cL[\ketbra{\psi_i}]+dx \cL'[\ketbra{\psi_i}] \right)\right) \nonumber\\
&\le& \sum_i p_i\dB^2\left(\cL[\ketbra{\psi_i}],
\cL[\ketbra{\psi_i}]+dx \cL'[\ketbra{\psi_i}] \right)\nonumber\\
&\le &\dB^2\left(\cL[\ketbra{\psi_0}],
\cL[\ketbra{\psi_0}]+dx \cL'[\ketbra{\psi_0}] \right)\,,
\end{eqnarray}
where we have designated $|\psi_0\rangle$ to be the state which maximizes $\dB^2\left(\cL[\ketbra{\psi_i}],
\cL[\ketbra{\psi_i}]+dx \cL'[\ketbra{\psi_i}] \right)$ over all
$|\psi_i\rangle$, and thus minimizes $\delta x_{\rm
  min}\left.\right|_{\cL[\ketbra{\psi_i}]}$. Plugging this upper bound on $\dB^2$ back
into 
(\ref{dxmin}) and taking the square root immediately shows the
theorem.
\end{proof}

Since there are many ways of mixing a given density matrix from pure
states, the usefulness of the theorem for obtaining bounds on the
best sensitivity achievable may depend on the chosen decomposition of
$\rho_0$.  For 
example, one may start with a pure 
state that is entangled over all subsystems, and admix to it
a separable 
state (see e.g.\cite{Werner89}).  Then the lower bound on $\delta x_{\rm min}$
remains given by the entangled state.  However, for a sufficiently large
admixture of the separable state, the initially entangled state might become
at least partially separable,  and a new decomposition of $\rho(x)$
in terms of pure states can be found, which raises the lower bound on
$\delta x_{\rm min}$. The most straightforward and useful application of the
theorem arises in the case of an initially fully separable state
$\rho_0$. In this case the theorem immediately shows that one cannot achieve
a better sensitivity than with the initial pure product states from which it
is mixed, i.e.~in particular a scaling as $1/\sqrt{K}$ with the number
of quantum resources $K$ for the example mentioned in the
introduction. This generalizes the findings of 
\cite{Giovannetti06} for the CC and CQ scenarios (classical initial
state, taken as a pure product state in \cite{Giovannetti06}, and
factorizing or entangling measurement) to mixed separable states.  \\

Very recently a preprint appeared which came to the same conclusion using a
different method, and in a
less general setting \cite{Goldstein10}.

\bibliography{../mybibs_bt}

\end{document}